\documentclass[fleqn,usenatbib]{mnras}

\usepackage{newtxtext,newtxmath}

\usepackage[T1]{fontenc}
\usepackage{ae,aecompl}

\usepackage{graphicx}	
\usepackage{amsmath}	
\usepackage{amssymb}	

\title[Telescope aperture, human vision, and sunspot counts]{The effect of telescope aperture, scattered 
light, and human vision on early measurements of sunspot and group 
numbers}

\author[N. V. Karachik et al.]{
Nina V. Karachik,$^{1}$\thanks{E-mail: ninakarachik@mail.ru}
Alexei  A. Pevtsov,$^{2}$\thanks{E-mail: apevtsov@nso.edu}
Yury A. Nagovitsyn$^{3,4}$\thanks{E-mail: nag@gaoran.ru}
\\
$^{1}$Ulugh Beg Astronomical Institute of the Uzbekistan Academy of Sciences, Tashkent, 100052, Uzbekistan\\
$^{2}$National Solar Observatory, Boulder, CO 80303, USA\\
$^{3}$Central Astronomical Observatory
of the Russian Academy of Sciences at Pulkovo,
St. Petersburg, 196140, Russia\\
$^{4}$St. Petersburg State University of Aerospace Instrumentation,
Bol'shaya Morskaya ul. 67, St. Petersburg, 190000, Russia
}

\date{Accepted XXX. Received YYY; in original form ZZZ}

\pubyear{2019}

\begin{document}
\label{firstpage}
\pagerange{\pageref{firstpage}--\pageref{lastpage}}
\maketitle

\begin{abstract}
Early telescopic observations of sunspots were conducted with instruments of
relatively small aperture. These instruments also suffered from 
a higher level of scattered light, and the human 
eye served as a ``detector''. The eye's ability 
to resolve small details depends on image contrast, and on 
average the intensity variations smaller than $\approx$ 3\% contrast relative to background are not
detected even if they are resolved by the telescope. Here we study the effect of these three parameters (telescope 
aperture, scattered light, and detection threshold of human vision) 
on sunspot number, group number, and area of sunspots.
As an ``ideal'' dataset, we employ white-light (pseudo-continuum)
observations from Helioseismic and Magnetic Imager (HMI) onboard of Solar Dynamics Observatory,
and we model the appearance of sunspots by degrading the HMI 
images to corresponding 
telescope apertures with an added scattered light. We discuss the
effects of different parameters on sunspot counts and derive 
functional dependencies, which could be used to normalize 
historical observations of sunspot counts to common denominator.
\end{abstract}




\section{Introduction}\label{sec:intro} 

Sunspot number is the longest time series representing the direct 
measurements of solar activity. The earliest telescope observations 
of sunspots go back to early 1610, when several scientists including 
Galileo Galilei, Thomas Harriot, Christoph Scheiner, and Johannes Fabricius
\citep{Hockey.etal2014} 
observed dark areas of irregular shape now known as 
sunspots. (The first publication about sunspot 
observations was by Johannes Fabricius in 1611, in his 
``Narratio de maculis in sole 
observatis 
et apparente earum cum sole conversione'').
After their discovery, the 
sunspots were observed by several scientists although some of them 
were searching for hypothesized planets orbiting the Sun and thus, 
their records could be biased towards dark features of a regular (round) shape (\citet{Zolotova.Ponyavin2015}, however, see \citet{Usoskin.etal2015,Carrasco.etal2018} for opposing views).
The later analysis of the historical records revealed that between
1645 and 1715 the sunspot sightings were extremely rare, and in fact,
that period of prolong sunspot minimum was christened the Maunder (Grand)
Minimum \citep{Eddy1976}. Despite lack of sunspots, the cosmogenic isotope records suggest that the magnetic cycles 
on the Sun did continue during the Maunder minimum 
\citep[e.g.,][]{Beer.etal1998,Cliver.etal1998,Poluianov.etal2014}, 
and one of the 
speculations put forward was that, perhaps, during that time period
the activity was represented by small sunspots, which were
hard to observe with the existing telescopes.
In fact, \citet{Vaquero.etal2015} 
found that the solar cycle activity 
did continue during Maunder minimum, but with a low amplitudes of less than 
5--10 Sunspot Number (SSN).

The return of sunspot activity was immediately noted by observers 
around the globe (see, for example, a letter to the Russian Czar Peter 
the Great by 
James Bruce, a Russian statesman and scientist known in Russia as 
Yakov Brus \citep{Hockey.etal2014}. In this letter, 
dated 18 July 1716, Bruce writes that he observed a great number of 
spots on the Sun and adds that to his knowledge, sunspots were not 
seeing for a long time). 
In addition to a very low number of sunspots during the 
Maunder Minimum, most of them were located in one (southern) 
hemisphere \citep[e.g.,][]{Ribes.Nesme-Ribes1993}. 
The hemispheric asymmetry in sunspot formation and their very small 
number during the Maunder Minimum have been interpreted
as a change in the character of solar activity. The reader 
can find a review of solar activity during Maunder minimum in \citet{Usoskin.etal2015}.

In addition, there could be
the effects of instrumentation (telescopes) used for sunspot observations.
In the past, the effect of telescope aperture on sunspot observations
was largely discarded on somewhat general statements that the telescope
apertures were sufficiently large to resolve even smallest pores 
\citep[e.g.,][]{Hoyt.Schatten1996}. However, this argument ignores the 
fact that many historical observers were stopping down the aperture of their 
telescopes
using aperture masks. This was usually done to reduce the brightness of solar image
and mitigate the heat load in prime focus. Thus, even if the telescopes 
used for astronomical observations had sufficiently large apertures during that
historical period, solar observations were made with significantly smaller
effective apertures. Furthermore, early observations of sunspots were eye observations,
and additional glass filters (gray or green) were used to further reduce 
the intensity of the sunlight. In combination with imperfections of
optical instruments of that time (e.g., spherical and chromatic aberrations) this will increase the level of the scattered light
in the telescope, thus, potentially affecting the detectability of small sunspots.
The effect of spherical and chromatic aberrations in sunspot observations in 18th century was considered by \citet[][]{Svalgaard2016,Svalgaard2017}, who set up an observing network of four amateur astronomers using original telescopes from the 18th Century with the same defects as the instruments available to observers of that period. Initial results found about one third of sunspot groups as compared with modern instruments.

Visual observations are also the subject of a physiological limitations of human eye -- a property that is rarely considered in the framework of astronomical observations \citep[but see pioneering work by ][]{Schaefer1991,Schaefer1993}. In this paper, we investigate these effects on sunspot and group numbers, and sunspot areas. In Section \ref{sec:method} we describe our approach to
modeling the effect of telescope aperture, scattered light, and human vision on sunspot and group number. 
Section \ref{sec:ssn} presents the results of modeling, and in Section 
\ref{sec:discuss}, we discuss our findings.

\section{Method and data}\label{sec:method}

The detectability of sunspots depends on a resolving power of a telescope, observing 
conditions (atmospheric seeing), and the sensitivity of a detector (a human eye in historical 
observations). These are the three primary effects that we include in our investigation.

Angular resolution $\theta$ of an ideal telescope of diameter (aperture) $\rm D$ in a monochromatic
light of wavelength $\lambda$ is determined by the Rayleigh criterion

\begin{equation}
    \theta = 1.22{\frac{\lambda}{D}}
    \label{eq:resolution}
\end{equation}

Taking as an example, $\rm D$=0.1~m (or about 4 inches) and $\lambda$=555~nm (maximum daylight 
sensitivity of a normal sighted human eye), we arrive to $\theta \approx$ 1.40 seconds of arc. 
Assuming that solar pores vary in diameter between 1 and 5 arcseconds \citep{Bruzek.Durrant1977},
one would require a telescope with an aperture of 0.14~m (for 1 arcsec resolution) 
and 0.028~m (for 5 arcseonds resolution) to
fully resolve pores of this size. Thus, the full aperture of telescopes used for astronomical observations during 1620--1675 
\citep[see Table 1 in ][]{Racine2004} should be sufficient to resolve large pores, while (large 
aperture) telescopes used after 1686 were capable to resolve even smallest pores.

After the invention of the telescope, their development for astronomical purposes was driven by 
a few individuals. Figure 1 in \citet{Racine2004} shows a break in improvements of telescopes closer 
to the end of Maunder minimum period. The article also refers to King's comment that ``the success of 
the long telescopes of the seventeenth century was due, very largely, to the painstaking and 
persistent efforts of men like Hevelius, Huygens and J. D. Cassini. Indeed, after Cassini's death in 
1712, his successors were unable to see what he had already discovered, let alone add to the list, 
and the telescopes gradually fell into disuse'' \citep[][page 133]{King1955,Racine2004}. 
Description of observations from that period of time often omits the details about the telescope 
aperture. \citet[][see, Section 3.3]{Ribes.Nesme-Ribes1993} provide an indirect reference for 
instruments of that time that ``usually a six-foot telescope would have an aperture of two inches and 
seven lines'', or, according to our calculations, about 0.065~m. As an additional complication, for solar observations the telescope apertures
were often stopped down (the aperture was reduced by using a circular 
diaphragm in front of a telescope). This was done to reduce the amount of
light in focal plane to prevent damage to focal 
lens or to the observer's eye. Restricting the aperture by the means of 
diaphragm changes the resolution of the telescope, but unfortunately, while 
the fact of stopping down the diameter of telescope is mentioned in some 
historical records, the effective aperture is not provided. Based on authors' personal experience with amateur astronomer  observations
of the Sun, prior to development of 
full aperture filters it was not uncommon to use the entrance diaphragm of 
40--50 mm (about 1.57 -- 1.97 inch) in diameter.

Atmospheric seeing conditions, optical quality of telescopes, and use of glass 
filters to further decrease the intensity of light in the focal plane had negative effects on sunspot 
observations too. Prior to 1733, objectives for all refractors employed a single lens design and were a 
subject of strong chromatic aberration. The chromatic aberration would result in a slight blurring of an image in a 
focal plane (due to a lateral shift of images in different wavelengths). In addition, optical 
aberrations and use of ``neutral density'' filters increased the amount of scattered light, and thus,
reduced the contrast of images. This could have a major effect on detectability of small sunspots due to some physiological limitations of a human eye.

The ability of a human eye to detected the brightness variations is the subject of both 
spatial 
resolution and the image contrast \citep[e.g.,][and references therein]{Carroll.Wiederman2014}. On average, for a human eye it is hard to detect the contrast variations less than about 3\%. Figure 1 
in \citet{Carroll.Wiederman2014} provides an example of an intensity pattern used for testing human vision. It displays 
a 
periodic pattern of intensity variation with different spatial 
frequencies (in the horizontal direction) and with different image 
contrast (in the vertical direction). The frequency patterns are 
well-resolved at high contrast, but as contrast decreases, the 
patterns at high and low spatial frequencies gradually disappear. 

The scattered light in the telescope (due to optical distortions, misalignment and filters) decreases
the image contrast, and thus, it would reduce the detectability of pores and small sunspots. So far, 
this effect was not taken into consideration in computation of sunspot number.

To model the effect of telescope aperture and image contrast on sunspot time series, we use the modern white-light 
(a pseudo-continuum) images from the Helioseismic and Magnetic Imager 
\citep[HMI, ][]{Scherrer.etal2012} on board Solar Dynamics Observatory (SDO). We treat the HMI white-light image as the reference representing true distribution of sunspots on 
solar surface. To mimic the appearance of observations with the telescope of a different 
aperture, we convolve the HMI images with a point-spread function 
of an appropriate width. 
For simplicity, we use a symmetric 2D Gaussian function for that matter. 
The Rayleigh criterion represents the angular separation between 
two point sources equal to the radius of the 
Airy disk. FWHM of Airy disk $FWHM_{Airy} = 1.024 \lambda D$, where $\lambda$ is the wavelength and $D$ is the telescope 
aperture. Gaussian function provides a good approximation of intensity distribution in Airy disk, 
and thus, in this work we employ a normalized FWHM of 
Gaussian function to represent the telescope aperture. $FWHM_{Gauss} = 2\sigma \sqrt{2 \ln 2}$, where 
$\sigma$ is the standard deviation.
For each image convolved with the Gaussian function, we 
apply intensity threshold to identify sunspots and pores, and we group the identified 
features based on their separation in latitude and longitude. According to 
our definition, to be considered as a group, 
sunspots should be located within 15 degrees in longitude and 7 degrees in latitude 
from the leading feature (spot or pore). These criteria are based on \citet{Tlatova.etal2018}.
The grouping is done by starting from the largest 
spot on each image, and identifying all spots that satisfy distance criteria to be counted 
as a group. The spots that are identified on this step as a group are excluded, and the 
process is repeated until all spots on the solar disk are identified 
with their respective groups. 

For each HMI image, we 
compute the sunspot number as 
\begin{equation}
SSN = k~(10~GN + N_S),
\label{eq:ssn}
\end{equation}
where $GN$ is number of 
groups (or group number), 
and $N_S$ is the number of spots for this image. Classical 
formula for SSN includes additional scaling $k$-coefficient,
which is used as a normalization between different observers.
Here we assume $k$ = 1.
The group and sunspot number identification is 
repeated for the same ``ideal telescope'' image degraded to represent observations with different telescope apertures.
We note that in this analysis we did not try to develop an algorithm for a perfect 
identification of sunspots and groups. The goal is to have a reasonable approach and apply 
such approach consistently to images representing telescopes of different aperture to see 
the overall tendency for changes in sunspot and group number due to telescope aperture.
To mimic the effect of a scattered light, we convolved the images with a scattering function representing 2\% or 5\% (two different levels of 
scattered light) of mean intensity of solar image. The resulting images were
used to identify sunspots. This level of a scattered light is consistent (if not low) with the modern telescope observations of a Mercury transit \citep[e.g.,][]{Briand.etal2006}.

To model the effect of human vision, when 
identifying the sunspots, we applied a 3\% contrast criterion. Intensity 
variations less than 3\% were not counted as sunspots.

\section{Computation of sunspot and group numbers}\label{sec:ssn}

\begin{figure} 
 \centerline{\hbox{\includegraphics[width=0.57\columnwidth,clip=]{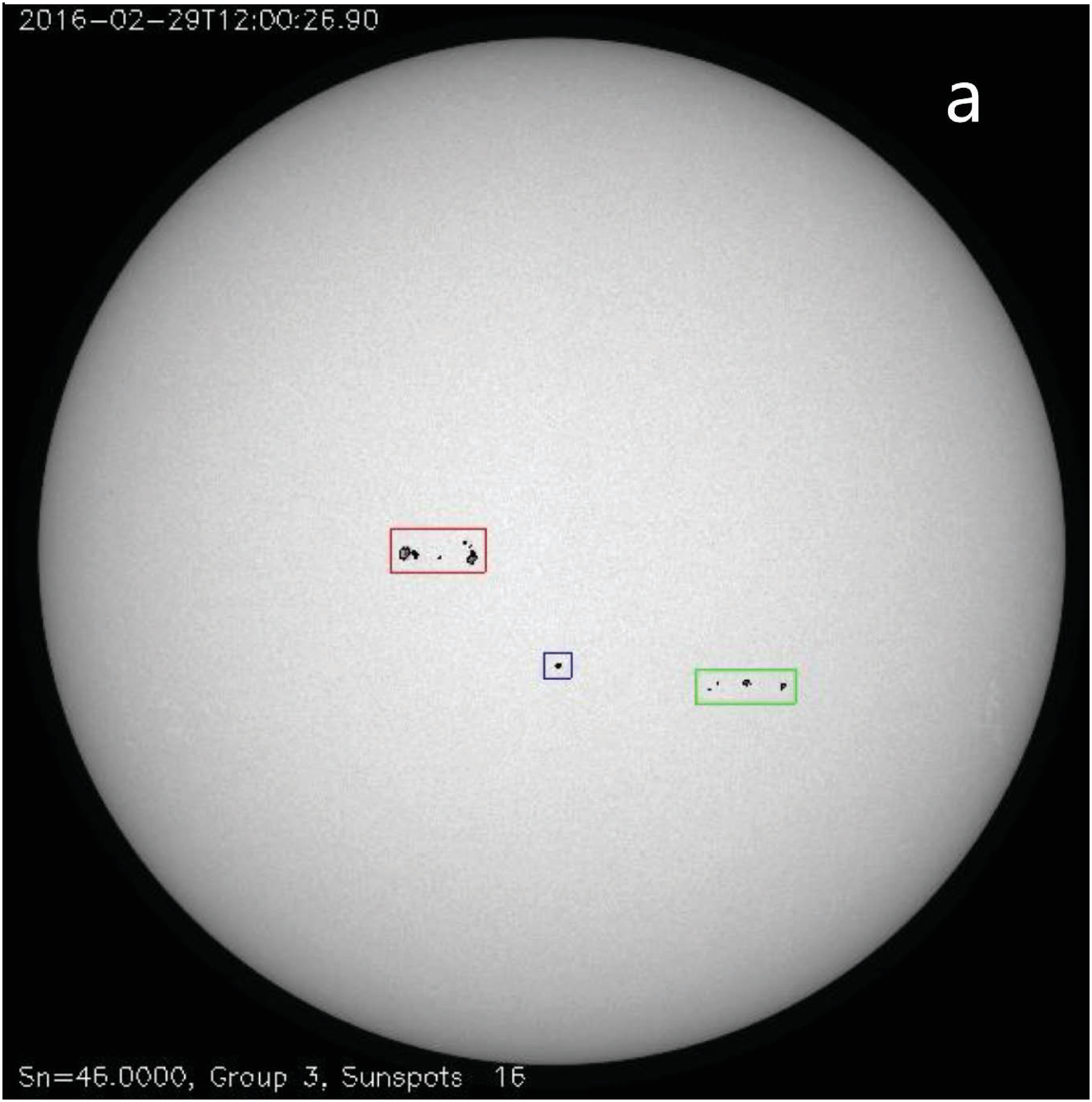}
 \vbox{\hbox{\includegraphics[width=0.43\columnwidth,clip=]{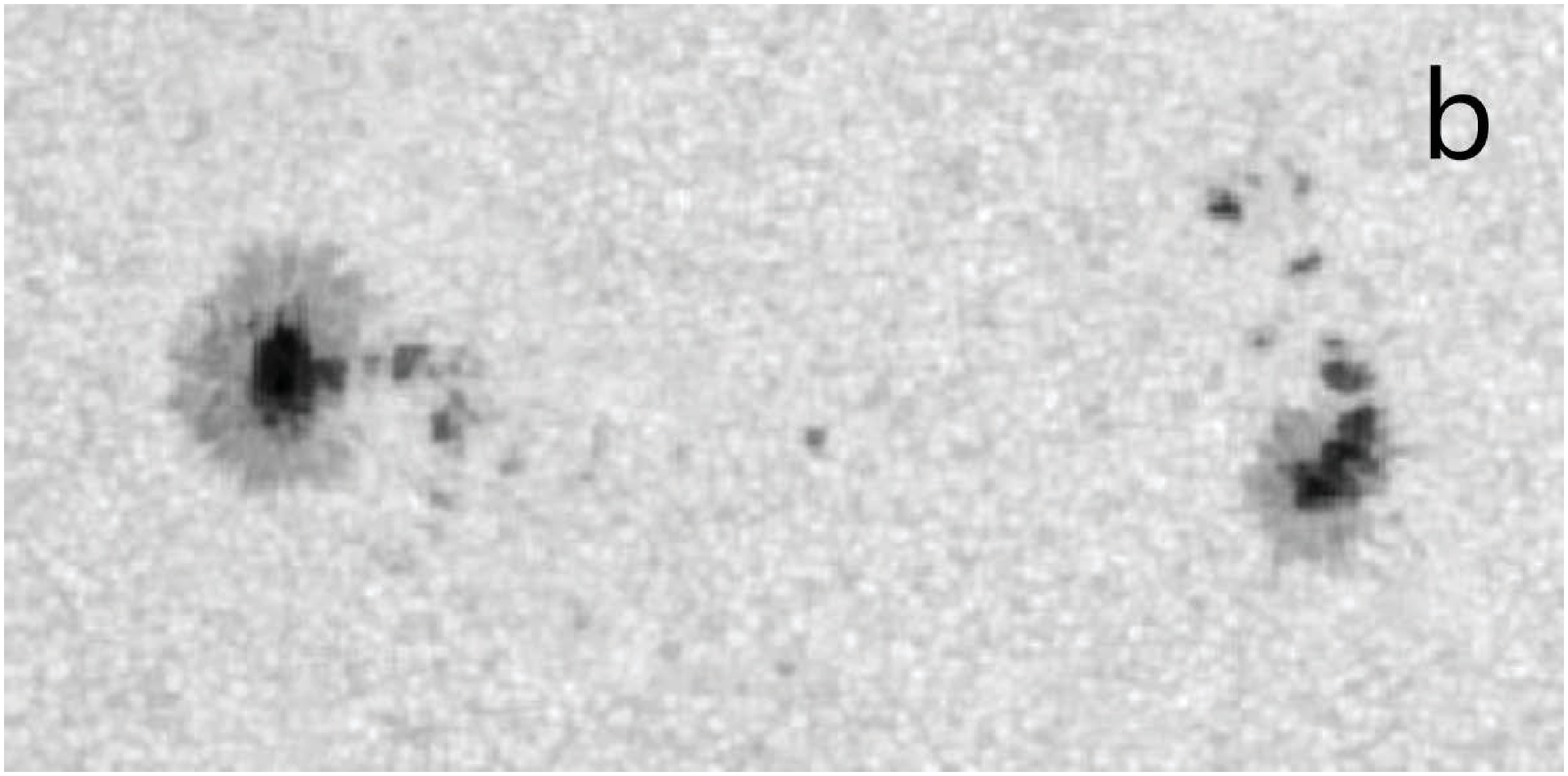}}
\hbox{\includegraphics[width=0.43\columnwidth,clip=]{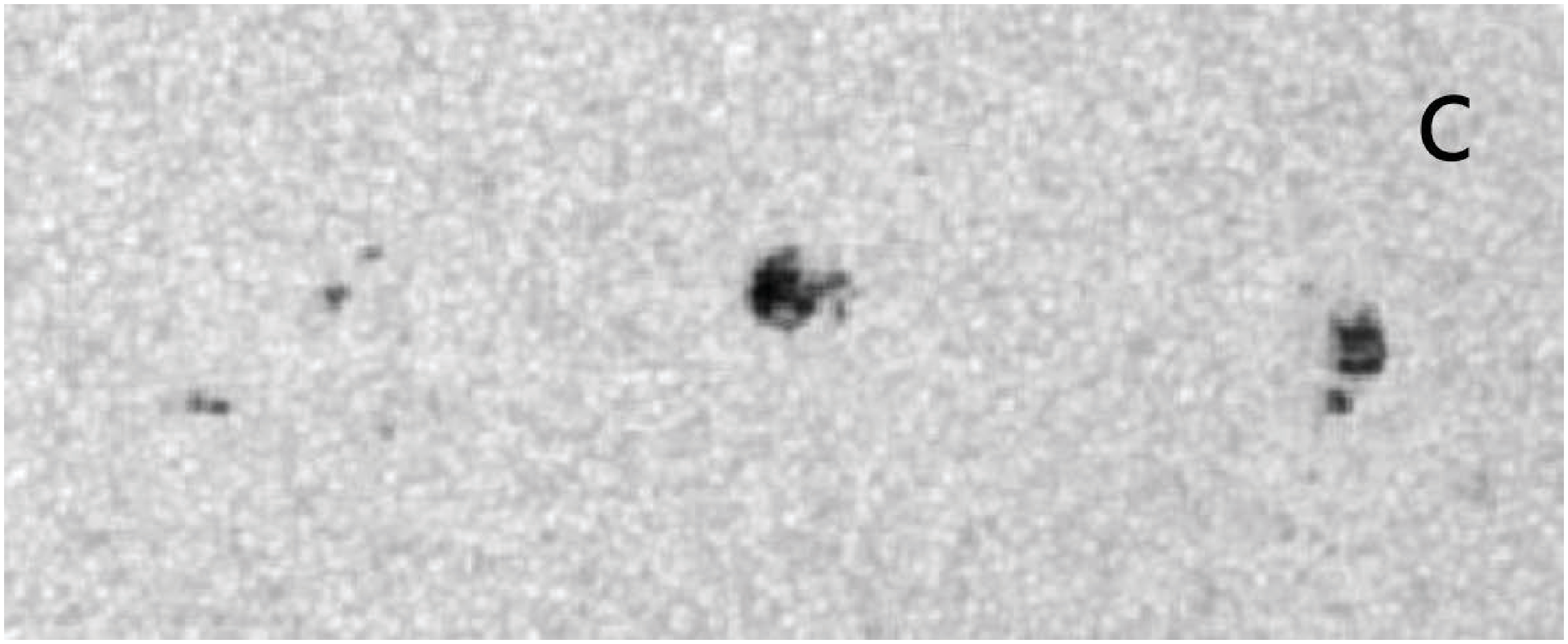}}
 \hbox{\includegraphics[width=0.2\columnwidth,clip=]{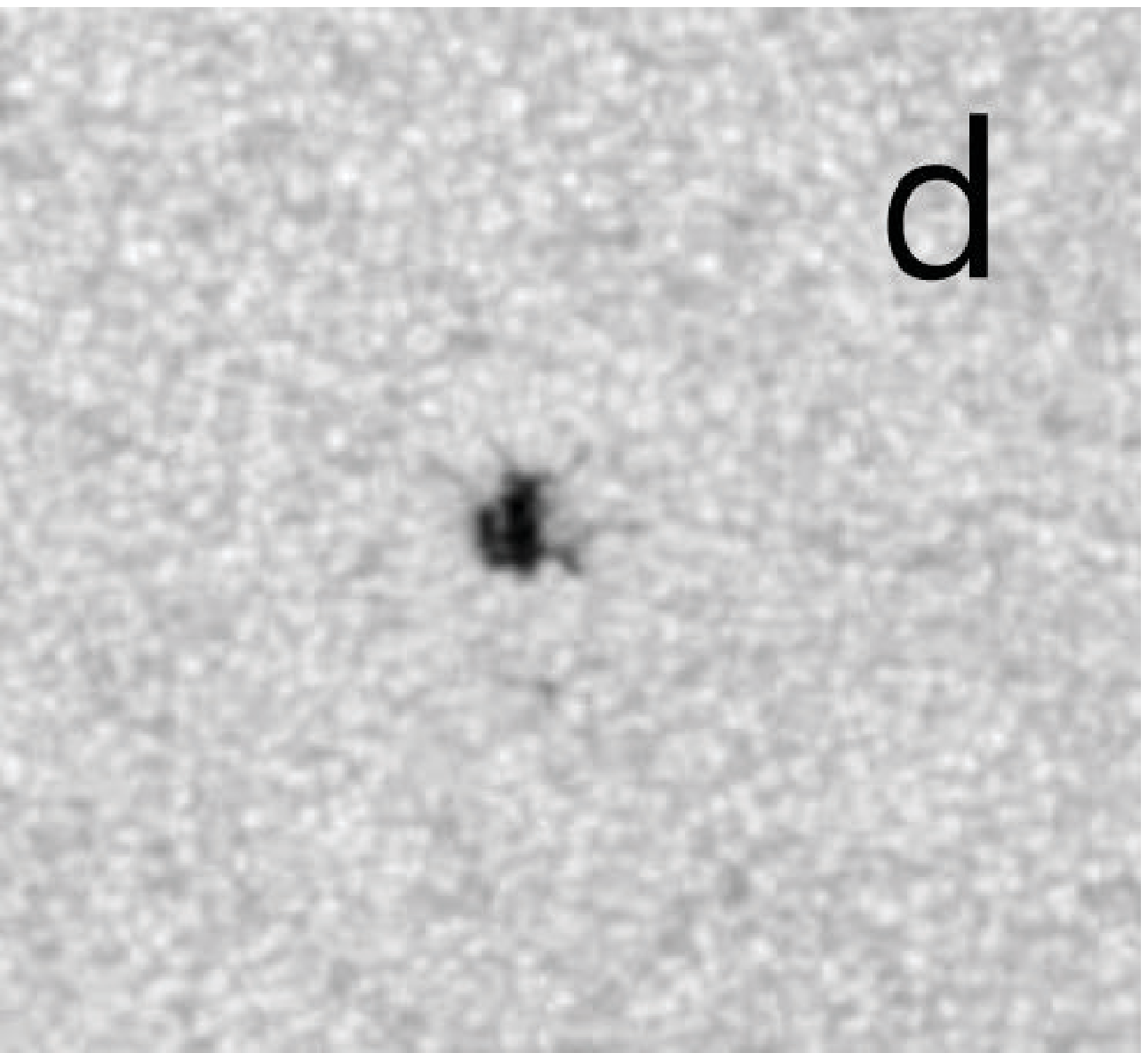}}
 }}}
 \caption{Example of full disk HMI image observed on 29 Feb. 2016 at 00:26 UT (a). Three colored boxes mark location of groups
 selected by our algorithm. Adjacent panels show magnifying view of each group. For group in panel (b), the algorithm identified 9 sunspots, panel (c) -- 6 sunspots, and panel (d) one spot.}

 \label{fig:hmi_ssn}
 \end{figure}

\begin{figure} 
\centerline{\includegraphics[width=1.0\columnwidth,clip=]{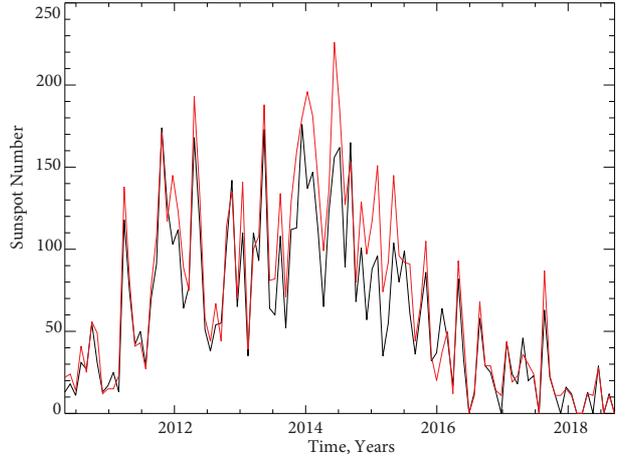}}
 \caption{Monthly international sunspot number from SILSO (black) and computed by our algorithm (red).}
 \label{fig:issn}
 \end{figure}

Figure \ref{fig:hmi_ssn} shows example of an HMI image with groups identified by our algorithm. Sunspot number (SSN) 
computed by our algorithm for this image is 46 (3 groups and 16 sunspots).
For comparison, the sunspot number from Sunspot Index and Long-term Solar Observations (SILSO) World Data Center lists 
SSN=45 for the same day. Figure \ref{fig:issn} shows international sunspot 
number time series and computed by us using full disk images from HMI/SDO.
In this comparison, only one daily measurement per month was used.
Two time series are in good agreement (Pearson correlation coefficient,
r$_P$=0.963), both showing the solar cycle variation for sunspot cycle 24. Other time series that we compared our 
SSN determination with is Sunspot Tracking and Recognition Algorithm \citep[STARA][]{Watson.etal2009}. The latter used
white-light observations from the Michelson Doppler Imager (MDI) on board Solar and Heliospheric Observatory (SOHO). Daily (one image per month) sunspot number determined by our algorithm also strongly correlate with STARA SSN (r$_P$=0.957) although our SSN tend to show a slightly lower values in (local) minima of time series. We explain this
by lower spatial resolution of the original datasets used by STARA. This could be related to a difference in a pixel size of images from two datasets. We use HMI data with pixel size of about 0.5 arcseconds, while STARA 
employed MDI/SOHO data with pixel size of about 2 arcseconds. Based on this comparison with two independent (SILSO and STARA) datasets,
we conclude that our algorithm performs sufficiently well for identifying the individual sunspots and sunspot groups on HMI 
white light images.

As the next step, we repeated our identification of sunspots for degraded HMI images mimicking observations with lower spatial
resolution and with added scattered light. As expected, the sunspot number decreases with decreasing telescope aperture. For example,
using observations shown in Figure \ref{fig:hmi_ssn}a degraded to 2 inch 
telescope (without scattered light) returns SSN = 39 (3 groups and 9 
sunspots). Table \ref{tab:difap} shows SSN and GN for other apertures 
corresponding to this image 
including naked eye (0.13 inch or about 3 mm).

Figures \ref{fig:ssn} - \ref{fig:ssn-area} show
Sunspot Number (SSN), Group Number (GN) and area of sunspots
as function of the telescope aperture. 
The data are normalized to the maximum of each parameter as observed by the telescope with 130 mm (5.12 inch) aperture and in the absence of scattered light. The plots include all sunspots and groups identified by our method in solar cycle 24. The data points shown in Figures \ref{fig:ssn}--\ref{fig:ssn-area} are also provided in Table \ref{tab:alldata}. Coefficients of fitted curves are listed in Table \ref{tab:logsq}.
In comparison with 130 mm aperture, an observer equipped with a 40 mm aperture telescope will measure about 60\% SSN, 78\% GN,  and 90\% sunspot areas. Having 5\% of scattered light will further reduce these values to 
55\%, 75\%, and 86\%, accordingly. The decrease in telescope aperture beyond 40 mm results in a rapid decrease in SSN, GN, and sunspot area. Out of three 
parameters, SSN number shows stronger dependence on the 
level of scattered light (Figure \ref{fig:ssn}), and sunspot areas are much less affected (Figure \ref{fig:ssn-area}).
This agrees with \citet{Nagovitsyn.Georgieva2017} findings that sunspot area is a more robust measure of sunspot activity as compared with SSN and GN.
The scattered light has a stronger effect on telescopes with larger apertures. Thus, for example, adding 5\% scattered light to 130 mm aperture telescope leads to
about 20\% reduction in SSN (Figure \ref{fig:ssn}) and 10\% reduction in GN (Figure \ref{fig:spots-groups}). The effect, however, is much smaller for small apertures. We see this as indication that for small apertures, the diffraction 
(telescope resolution) has much stronger effect as compared with
the scattered light, and for larger apertures, the latter becomes
more important.
In the handbook for amateur solar observations, 
\citet{Beck1995} recommend on optimal telescope aperture of 80 mm. This 
recommendation is based on a typical diameter of near-ground turbulent 
convection cells, and the resolving power necessary to see 
the smallest 
spots. Telescopes with larger aperture will have an image blurred 
(and hence, less contrasted) due to the light passing through 
multiple convection cells as compared with 80 mm aperture scope, 
which images will be affected by a single cell. The results shown 
in Figures \ref{fig:ssn}--\ref{fig:ssn-area} indicate that a telescope with such 
aperture performs sufficiently well even in presence of the scattered light although
our current calculations do not take into account the effects of the atmospheric seeing.

\section{Discussion}\label{sec:discuss}

Our results clearly demonstrate how the telescope aperture, its optical quality 
(scattered light) in combination with physiological limitations of human vision can affect the detectability of sunspots. We think that previous, somewhat generic, 
claims that during the Maunder minimum the telescopes had sufficiently high apertures 
to resolve ''regular diameter'' sunspots may ignore the fact that for solar observations
the entrance aperture of telescopes was 
usually stopped down using a diaphragm of a 
much
smaller diameter. The latter would result in 
reducing the resolving power of an instrument.
While the historical records do mention 
stopping down the telescope aperture, no
aperture of the diaphragm is usually provided.
Moreover, the early telescopes were not perfect in respect to their optical properties and thus, may have had a significant level of
scattered light.

There are some historical records that seem to support the notion that the detectability of sunspots 
could be severely affected by the observational conditions (e.g., the instrument quality,
the atmospheric seeing conditions and the observer's vision). For example, Table 1 and Figure 1 in \citet{Vaquero.etal2007} provides sunspot records for the observations taken 
simultaneously by different observers. For June 3, 1769 observations, number of groups 
(GN) detected by different observers vary between 1 and 10. Even for observers in 
the same geographic location (e.g., Paris), the sunspot group number vary significantly: GN=1 
(Darquier), GN=5 (Bailly), and GN=9 (Messier). Based on \citet{Tisserand1881}, 
Darquier probably used a small aperture instrument (a quadrant with 27 inch focal length telescope). According to \citet{Delalande.Messier1769} letters, Messier used achromatic telescope (refractor) with 12 feet focus, 3.75 inch aperture and 180 magnification, 
while Bailly was using a reflector of 30 inch focus and 4.5 inch aperture. Messier also 
mentions that seeing conditions were not great (``vapors'' and clouds).  This seems to 
agree with our conjecture that observing with small aperture telescope (Darquier) returns small 
number of groups, and using a telescope with less scattered light (Messier) allows 
detecting a larger number of groups as compared with a telescope of similar aperture 
but higher level of scattered light (Bailly).

The detectability of sunspots could also be 
affected by difference in persons' vision, 
which is hard, if possible to characterise 
without additional information. 
As one example, we use an interesting 
handwritten note on the sunspot drawing taken 
at Mount Wilson Observatory (MWO) on 15 
February 1999 (scanned image available via 
UCLA server at \url{ftp://howard.astro.ucla.edu/pub/obs/drawings} does not contain this note).
This day, two naked eye sunspots were present on the Sun. However, the note on the drawing indicates that out of three
observers, L.W. (Larry Webster) saw 2 naked eye sunspots, P.G. (Peter Gilman) 
saw only one (lower) sunspot, and S.P. (Steve Padilla) 
saw none of naked eye sunspots. These were the trained professional 
observers, who probably knew about the presence of naked-eye 
sunspots on solar disk the day of observations, but still 
each would see them differently. While the naked eye sunspots are not
the subject of our  paper, we use this as an example of how 
human vision could affect the detectability of sunspots. 

The atmospheric seeing is other 
unknown parameter that could affected the detectalibity of sunspots. For historical observations, it is hard if possible to estimate, but for a pictorial example, we 
refer the reader to sunspot drawings made at MWO over three consecutive days 15-17 December 1969. In the first drawing (15 Dec. 1969),
taken under almost excellent atmospheric seeing conditions (seeing = 4+ on a scale 1--5), the observer identified 9 different groups with large number of small sunspots and pores. Second image was taken on 16 Dec. 1969
 under extremely poor conditions (seeing = 1) and
 only 4 groups consisting large spots and a few pores were identified.
 Observations taken on 17 Dec. 1969 under good seeing~ =~3 conditions again show 10 groups with  a large number of small pores.

Figure \ref{fig:cycle24} shows sunspot number (SSN) computed for cycle 24 as if sunspots were observed with the small aperture telescopes and 5\% scattered light. For a reference, Galileo used 0.5 inch telescope for his early sunspot observations. 
(The largest telescope that Galileo used had an aperture of 5.1 cm, but was usually stopped down to 2.6 cm, or about one inch. However, the tests of one of Galileo's first telescopes showed the resolving power corresponding to about 0.5 inch aperture (resolving $\approx$ 10 arcsec and larger features, \citet[][]{Arlt.etal2016}.)
Schwabe employed 1.25 inch instrument, and Wolf used 2.5 inch telescope. It is clear that naked eye observations would not allow detecting the solar cycle variations for cycle similar in amplitude to cycle 24. 
Galileo and Schwabe instruments would be marginal in detecting a cycle 
similar in amplitude to cycle 24 if one takes into account the scattered
light and 3\% detectability threshold of human eye. Observations with  
telescopes larger than 2 inch in aperture will detect 
the cycle variations albeit with a significantly smaller 
amplitude (about 50\% less according to  Figure 
\ref{fig:cycle24}, compare black and blue lines).

\begin{figure} 
\centerline{\includegraphics[width=1.0\columnwidth,clip=]{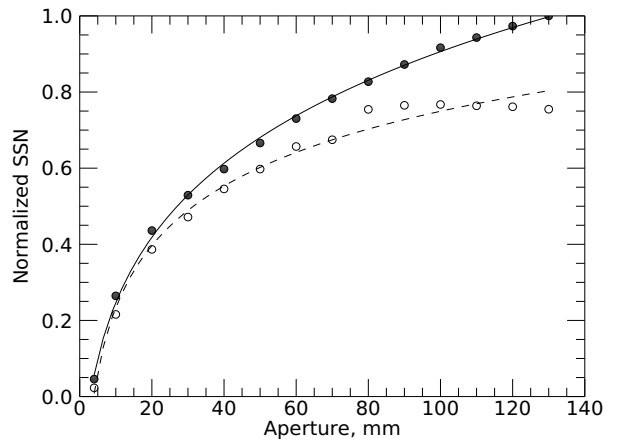}}
 \caption{Sunspot number (SSN) computed from HMI images as a function of effective aperture of a telescope without scattered light (filled circles) and with 5\% scattered light (open circles). Data are normalized to maximum value of SSN without scattered light. Solid lines show least-square fitting by logsquare function (see coefficients in Table \ref{tab:logsq}).}
 \label{fig:ssn}
 \end{figure}

\begin{figure} 
\centerline{\includegraphics[width=1.0\columnwidth,clip=]{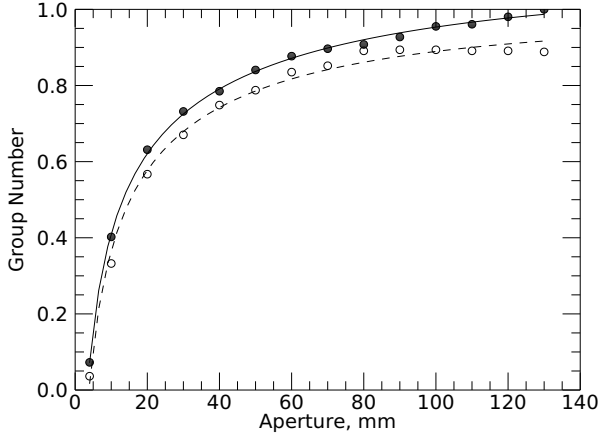}}
\caption{Same as Figure \ref{fig:ssn}, but for group number (GN).}
\label{fig:spots-groups}
 \end{figure}

\begin{figure} 
\centerline{\includegraphics[width=1.0\columnwidth,clip=]{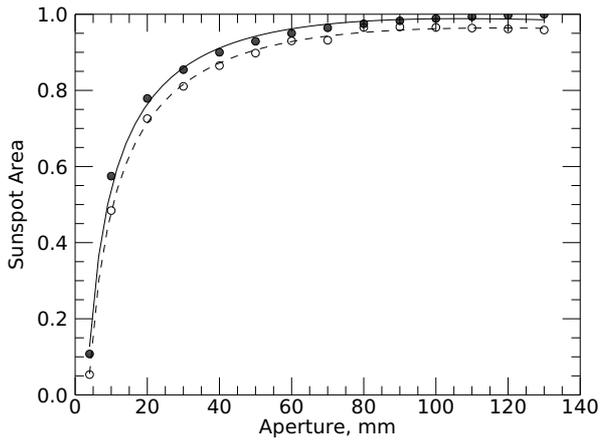}}
\caption{Same as Figure \ref{fig:ssn}, but for sunspot area.}
\label{fig:ssn-area}
 \end{figure}

\begin{figure} 
\centerline{\includegraphics[width=1.0\columnwidth,clip=]{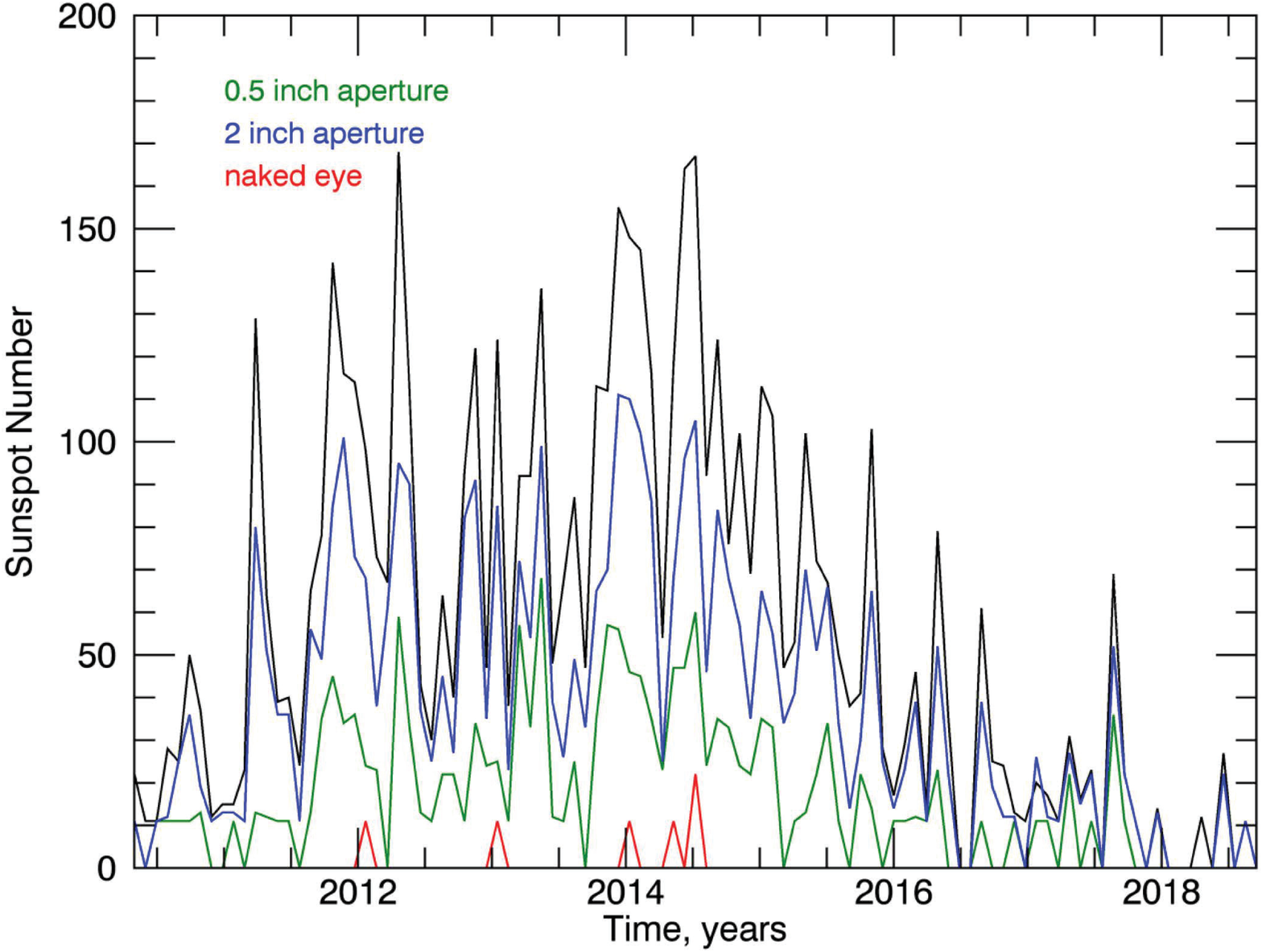}}
 \caption{Monthly sunspot number (SSN) in cycle 24 computed for telescopes of different aperture: full resolution (black), 2 inch aperture (blue), 0.5 inch (green) and naked eye observations (red).}
 \label{fig:cycle24}
 \end{figure}

While some effects of human vision on sunspot number the cannot be 
fully quantified (e.g., change in contrast detection sensitivity with 
age), our results offer a path for normalizing historical SSN time 
series to bring observations 
taken with the telescopes of different apertures and scattered light in line 
with each other. One interesting 
observation that the reader can note from Figure \ref{fig:ssn} and Table \ref{tab:alldata} is that for 
60 mm ($\approx$ 2.5 inch) telescope, SSN is about 60\% of 
``true'' SSN. Although it could be a coincidence, this is 
close to $k$-coefficient used by Wolf when computing 
sunspot number (see Equation \ref{eq:ssn}).

Our results indicate that sunspot number (SSN) 
is affected by the telescope quality (including scattered light), human 
vision, and the observing conditions more than group number and sunspot 
areas. In their turn, sunspot area is a much more robust measure than 
SSN and GN. However, while
sunspot areas could be derived even 
from the very early drawings, they 
might be a subject of large uncertainties due to different drawing styles, 
lack of proper scaling, and some may even be unsuitable for a scientific use \citep[e.g.,][]{Arlt.etal2016,Senthamizh.Pavai.etal2015,Carrasco.etal2018b,Fujiyama.etal2019}.

Many details of early telescopic 
observations are unknown including focal length, aperture, and other  
characteristics on the telescopes, the used methodology, and even the 
aims of these observations \citep[]{Munoz.Vaquero2019}. While we base
some of our interpretation on the assumption that all early observations were taken with small aperture telescopes, it is likely, that the telescopes of different (small, medium and large) apertures were used.
A similar mix of instruments was also used for recording sunspot numbers
in later periods (18th and 19th centuries). Thus, the results of our 
study could also be applicable to sunspot visual observations taken in other periods during last four centuries, not only to the earliest observations.


%
\begin{table}
\caption{Change in SSN as function of telescope aperture for image shown in Figure \ref{fig:hmi_ssn}}\label{tab:difap}
\begin{tabular}{rrrr}     
\hline
Aperture&GN&Spots&SSN\\
(inch)&&&\\
\hline
ideal&3&16&46\\
2.00&3&9&39\\
1.00&3&6&36\\
0.64&3&4&34\\
0.57&2&3&23\\
0.37&1&2&12\\
0.13&0&0&0\\
\hline
\end{tabular}
\end{table}

\begin{table}
\caption{Sunspot, Group number and sunspot area without scattered light and with 5\% scattered light}\label{tab:alldata}
\begin{tabular}{rrrrrrr}     
\hline
mm&SSN&SSN+S&Area&Area+S&GN&GN+S\\
\hline
4&0.046&0.023&0.108&0.053&0.073&0.036\\
  10&0.265&0.216&0.575&0.484&0.402&0.332\\
  20&0.436&0.387&0.779&0.726&0.631&0.567\\
  30&0.529&0.472&0.854&0.811&0.732&0.670\\
  40&0.598&0.545&0.900&0.865&0.785&0.749\\
  50&0.666&0.598&0.929&0.898&0.841&0.788\\
  60&0.730&0.657&0.950&0.930&0.877&0.835\\
  70&0.783&0.675&0.964&0.932&0.897&0.852\\
  80&0.827&0.755&0.975&0.965&0.908&0.891\\
  90&0.872&0.765&0.983&0.967&0.927&0.894\\
 100&0.917&0.767&0.989&0.965&0.955&0.894\\
 110&0.943&0.764&0.994&0.964&0.961&0.891\\
 120&0.973&0.761&0.997&0.961&0.980&0.891\\
 130&1.000&0.755&1.000&0.959&1.000&0.888\\
\hline
\end{tabular}
\end{table}

\begin{table}
\caption{Coefficients of $y=a_0+a_1\log(x)+a_2\log^2(x)$ function fitted to data shown in Figures \ref{fig:ssn}--\ref{fig:ssn-area}}\label{tab:logsq}
\begin{tabular}{rrrrrrr}     
\hline
&SSN&SSN+S&GN&GN+S&Area&Area+S\\
\hline
a$_0$&-0.153&-0.356&-0.576&-0.666&-0.754&-0.843\\
a$_1$&0.271&0.626&1.210&1.286&1.717&1.747\\
a$_2$&0.129&-0.037&-0.222&-0.254&-0.423&-0.422\\
\hline
\end{tabular}
\end{table}


\section*{Acknowledgements}

The authors thank Mr. Alexander Pevtsov for providing a 
copy 
of image of MWO sunspot drawings showing the notes on 
naked-eye observations on 15 Feb. 
1999 from the archives of the Carnegie Observatories in 
Pasadena, CA. We thank Dr. L. Svalgaard and other anonymous 
reviewer for 
their constructive comments, which helped improving the article.
A portion of this work was carried out by N.V.K. through the short-term 
scientific internship program supported by Ministry of Innovative Development 
of the Republic of Uzbekistan.
A.A.P. acknowledges NASA NNX15AE95G grant, and 
the international team on Recalibration of the Sunspot 
Number Series
supported by the International Space Science Institute (ISSI), 
Bern, Switzerland. Y.A.N. work was supported by the Russian Foundation for 
Basic Research via RFBR 19-02-00088 grant and K$\Pi$19-270 program of
the Ministry of Education and Science of the Russian Federation. 
The U.S. National Solar 
Observatory (NSO) is operated by the Association of Universities for 
Research in Astronomy (AURA), Inc., under cooperative agreement with 
the National Science Foundation.
We acknowledge using sunspot and group numbers from WDC-SILSO, Royal 
Observatory of Belgium, Brussels.




\bibliographystyle{mnras}
\bibliography{pevtsov_bibliography} 

\begin{thebibliography}{}
\makeatletter
\relax
\def\mn@urlcharsother{\let\do\@makeother \do\$\do\&\do\#\do\^\do\_\do\%\do\~}
\def\mn@doi{\begingroup\mn@urlcharsother \@ifnextchar [ {\mn@doi@}
  {\mn@doi@[]}}
\def\mn@doi@[#1]#2{\def\@tempa{#1}\ifx\@tempa\@empty \href
  {http://dx.doi.org/#2} {doi:#2}\else \href {http://dx.doi.org/#2} {#1}\fi
  \endgroup}
\def\mn@eprint#1#2{\mn@eprint@#1:#2::\@nil}
\def\mn@eprint@arXiv#1{\href {http://arxiv.org/abs/#1} {{\tt arXiv:#1}}}
\def\mn@eprint@dblp#1{\href {http://dblp.uni-trier.de/rec/bibtex/#1.xml}
  {dblp:#1}}
\def\mn@eprint@#1:#2:#3:#4\@nil{\def\@tempa {#1}\def\@tempb {#2}\def\@tempc
  {#3}\ifx \@tempc \@empty \let \@tempc \@tempb \let \@tempb \@tempa \fi \ifx
  \@tempb \@empty \def\@tempb {arXiv}\fi \@ifundefined
  {mn@eprint@\@tempb}{\@tempb:\@tempc}{\expandafter \expandafter \csname
  mn@eprint@\@tempb\endcsname \expandafter{\@tempc}}}

\bibitem[\protect\citeauthoryear{{Arlt}, {Senthamizh Pavai}, {Schmiel}  \&
  {Spada}}{{Arlt} et~al.}{2016}]{Arlt.etal2016}
{Arlt} R.,  {Senthamizh Pavai} V.,  {Schmiel} C.,   {Spada} F.,  2016, \mn@doi
  [\aap] {10.1051/0004-6361/201629000}, 595, A104

\bibitem[\protect\citeauthoryear{{Beck}, {Hilbrecht}, {Reinsch}  \&
  {Voelker}}{{Beck} et~al.}{1995}]{Beck1995}
{Beck} R.,  {Hilbrecht} H.,  {Reinsch} K.,   {Voelker} P.,  1995, {Solar
  astronomy handbook}.
Willmann-Bell

\bibitem[\protect\citeauthoryear{{Beer}, {Tobias}  \& {Weiss}}{{Beer}
  et~al.}{1998}]{Beer.etal1998}
{Beer} J.,  {Tobias} S.,   {Weiss} N.,  1998, \mn@doi [\solphys]
  {10.1023/A:1005026001784}, \href
  {http://adsabs.harvard.edu/abs/1998SoPh..181..237B} {181, 237}

\bibitem[\protect\citeauthoryear{{Briand}, {Mattig}, {Ceppatelli}  \&
  {Mainella}}{{Briand} et~al.}{2006}]{Briand.etal2006}
{Briand} C.,  {Mattig} W.,  {Ceppatelli} G.,   {Mainella} G.,  2006, \mn@doi
  [\solphys] {10.1007/s11207-006-0033-5}, 234, 187

\bibitem[\protect\citeauthoryear{{Bruzek} \& {Durrant}}{{Bruzek} \&
  {Durrant}}{1977}]{Bruzek.Durrant1977}
{Bruzek} A.,  {Durrant} C.~J.,  eds, 1977, {Illustrated glossary for solar and
  solar-terrestrial physics}  Astrophysics and Space Science Library Vol. 69,
  \mn@doi{10.1007/978-94-010-1245-4.
}

\bibitem[\protect\citeauthoryear{{Carrasco}, {Vaquero}  \&
  {Gallego}}{{Carrasco} et~al.}{2018a}]{Carrasco.etal2018}
{Carrasco} V.~M.~S.,  {Vaquero} J.~M.,   {Gallego} M.~C.,  2018a, \mn@doi
  [\solphys] {10.1007/s11207-018-1270-0}, \href
  {https://ui.adsabs.harvard.edu/abs/2018SoPh..293...51C} {293, 51}

\bibitem[\protect\citeauthoryear{{Carrasco}, {Vaquero}, {Arlt}  \&
  {Gallego}}{{Carrasco} et~al.}{2018b}]{Carrasco.etal2018b}
{Carrasco} V.~M.~S.,  {Vaquero} J.~M.,  {Arlt} R.,   {Gallego} M.~C.,  2018b,
  \mn@doi [\solphys] {10.1007/s11207-018-1322-5}, \href
  {https://ui.adsabs.harvard.edu/abs/2018SoPh..293..102C} {293, 102}

\bibitem[\protect\citeauthoryear{{Cliver}, {Boriakoff}  \& {Bounar}}{{Cliver}
  et~al.}{1998}]{Cliver.etal1998}
{Cliver} E.~W.,  {Boriakoff} V.,   {Bounar} K.~H.,  1998, \mn@doi [\grl]
  {10.1029/98GL00500}, \href
  {http://adsabs.harvard.edu/abs/1998GeoRL..25..897C} {25, 897}

\bibitem[\protect\citeauthoryear{{Eddy}}{{Eddy}}{1976}]{Eddy1976}
{Eddy} J.~A.,  1976, \mn@doi [Science] {10.1126/science.192.4245.1189}, \href
  {http://adsabs.harvard.edu/abs/1976Sci...192.1189E} {192, 1189}

\bibitem[\protect\citeauthoryear{{Fujiyama} et~al.,}{{Fujiyama}
  et~al.}{2019}]{Fujiyama.etal2019}
{Fujiyama} M.,  et~al., 2019, \mn@doi [\solphys] {10.1007/s11207-019-1429-3},
  \href {https://ui.adsabs.harvard.edu/abs/2019SoPh..294...43F} {294, 43}

\bibitem[\protect\citeauthoryear{{Hockey}, {Trimble}, {Williams}, {Bracher},
  {Jarrell}, {March{\'e}ll}, {Palmeri}  \& {Green}}{{Hockey}
  et~al.}{2014}]{Hockey.etal2014}
{Hockey} T.,  {Trimble} V.,  {Williams} T.~R.,  {Bracher} K.,  {Jarrell} R.~A.,
   {March{\'e}ll} J.~D.,  {Palmeri} J.,   {Green} D.~W.~E.,  eds, 2014,
  {Biographical Encyclopedia of Astronomers}, 2 edn.
Springer, New York, NY, \mn@doi{10.1007/978-1-4419-9917-7}

\bibitem[\protect\citeauthoryear{{Hoyt} \& {Schatten}}{{Hoyt} \&
  {Schatten}}{1996}]{Hoyt.Schatten1996}
{Hoyt} D.~V.,  {Schatten} K.~H.,  1996, \mn@doi [\solphys]
  {10.1007/BF00149097}, \href
  {http://adsabs.harvard.edu/abs/1996SoPh..165..181H} {165, 181}

\bibitem[\protect\citeauthoryear{{King}}{{King}}{1979}]{King1955}
{King} H.~C.,  1979, {The History of the Telescope}.
Dover Publications, New York

\bibitem[\protect\citeauthoryear{{Mu{\~n}oz-Jaramillo} \&
  {Vaquero}}{{Mu{\~n}oz-Jaramillo} \& {Vaquero}}{2019}]{Munoz.Vaquero2019}
{Mu{\~n}oz-Jaramillo} A.,  {Vaquero} J.~M.,  2019, \mn@doi [Nature Astronomy]
  {10.1038/s41550-018-0638-2}, \href
  {https://ui.adsabs.harvard.edu/abs/2019NatAs...3..205M} {3, 205}

\bibitem[\protect\citeauthoryear{{Nagovitsyn} \& {Georgieva}}{{Nagovitsyn} \&
  {Georgieva}}{2017}]{Nagovitsyn.Georgieva2017}
{Nagovitsyn} Y.~A.,  {Georgieva} K.,  2017, \mn@doi [Geomagnetism and Aeronomy]
  {10.1134/S0016793217070131}, 57, 783

\bibitem[\protect\citeauthoryear{{O{'}Carroll} \& {Wiederman}}{{O{'}Carroll} \&
  {Wiederman}}{2014}]{Carroll.Wiederman2014}
{O{'}Carroll} D.~C.,  {Wiederman} S.~D.,  2014, \mn@doi [Phil. Trans. R. Soc.
  B] {10.1098/rstb.2013.0043}, 369, 20130043

\bibitem[\protect\citeauthoryear{{Poluianov}, {Usoskin}  \&
  {Kovaltsov}}{{Poluianov} et~al.}{2014}]{Poluianov.etal2014}
{Poluianov} S.~V.,  {Usoskin} I.~G.,   {Kovaltsov} G.~A.,  2014, \mn@doi
  [\solphys] {10.1007/s11207-014-0587-6}, \href
  {http://adsabs.harvard.edu/abs/2014SoPh..289.4701P} {289, 4701}

\bibitem[\protect\citeauthoryear{{Racine}}{{Racine}}{2004}]{Racine2004}
{Racine} R.,  2004, \mn@doi [\pasp] {10.1086/380955}, \href
  {http://adsabs.harvard.edu/abs/2004PASP..116...77R} {116, 77}

\bibitem[\protect\citeauthoryear{{Ribes} \& {Nesme-Ribes}}{{Ribes} \&
  {Nesme-Ribes}}{1993}]{Ribes.Nesme-Ribes1993}
{Ribes} J.~C.,  {Nesme-Ribes} E.,  1993, \aap, 276, 549

\bibitem[\protect\citeauthoryear{{Schaefer}}{{Schaefer}}{1991}]{Schaefer1991}
{Schaefer} B.~E.,  1991, \qjras, \href
  {https://ui.adsabs.harvard.edu/abs/1991QJRAS..32...35S} {32, 35}

\bibitem[\protect\citeauthoryear{{Schaefer}}{{Schaefer}}{1993}]{Schaefer1993}
{Schaefer} B.~E.,  1993, \mn@doi [\apj] {10.1086/172895}, \href
  {https://ui.adsabs.harvard.edu/abs/1993ApJ...411..909S} {411, 909}

\bibitem[\protect\citeauthoryear{{Scherrer} et~al.,}{{Scherrer}
  et~al.}{2012}]{Scherrer.etal2012}
{Scherrer} P.~H.,  et~al., 2012, \mn@doi [\solphys]
  {10.1007/s11207-011-9834-2}, \href
  {http://adsabs.harvard.edu/abs/2012SoPh..275..207S} {275, 207}

\bibitem[\protect\citeauthoryear{{Senthamizh Pavai}, {Arlt}, {Dasi-Espuig},
  {Krivova}  \& {Solanki}}{{Senthamizh Pavai}
  et~al.}{2015}]{Senthamizh.Pavai.etal2015}
{Senthamizh Pavai} V.,  {Arlt} R.,  {Dasi-Espuig} M.,  {Krivova} N.~A.,
  {Solanki} S.~K.,  2015, \mn@doi [\aap] {10.1051/0004-6361/201527080}, \href
  {https://ui.adsabs.harvard.edu/abs/2015A&A...584A..73S} {584, A73}

\bibitem[\protect\citeauthoryear{{Svalgaard}}{{Svalgaard}}{2016}]{Svalgaard2016}
{Svalgaard} L.,  2016, in AAS/Solar Physics Division Abstracts \#47. AAS/Solar
  Physics Division Meeting.
p. 10.12

\bibitem[\protect\citeauthoryear{{Svalgaard}}{{Svalgaard}}{2017}]{Svalgaard2017}
{Svalgaard} L.,  2017, \mn@doi [\solphys] {10.1007/s11207-016-1023-x}, \href
  {https://ui.adsabs.harvard.edu/abs/2017SoPh..292....4S} {292, 4}

\bibitem[\protect\citeauthoryear{{Tisserand}}{{Tisserand}}{1881}]{Tisserand1881}
{Tisserand} F.,  1881, Astronomical register, \href
  {http://adsabs.harvard.edu/abs/1881AReg...19..218T} {19, 218}

\bibitem[\protect\citeauthoryear{{Tlatova} et~al.,}{{Tlatova}
  et~al.}{2018}]{Tlatova.etal2018}
{Tlatova} K.,  et~al., 2018, \mn@doi [\solphys] {10.1007/s11207-018-1337-y},
  \href {http://adsabs.harvard.edu/abs/2018SoPh..293..118T} {293, 118}

\bibitem[\protect\citeauthoryear{{Usoskin} et~al.,}{{Usoskin}
  et~al.}{2015}]{Usoskin.etal2015}
{Usoskin} I.~G.,  et~al., 2015, \mn@doi [\aap] {10.1051/0004-6361/201526652},
  \href {https://ui.adsabs.harvard.edu/abs/2015A&A...581A..95U} {581, A95}

\bibitem[\protect\citeauthoryear{{Vaquero}, {Trigo}, {Gallego}  \&
  {Moreno-Corral}}{{Vaquero} et~al.}{2007}]{Vaquero.etal2007}
{Vaquero} J.~M.,  {Trigo} R.~M.,  {Gallego} M.~C.,   {Moreno-Corral} M.~A.,
  2007, \mn@doi [\solphys] {10.1007/s11207-006-0264-5}, \href
  {http://adsabs.harvard.edu/abs/2007SoPh..240..165V} {240, 165}

\bibitem[\protect\citeauthoryear{{Vaquero}, {Kovaltsov}, {Usoskin}, {Carrasco}
  \& {Gallego}}{{Vaquero} et~al.}{2015}]{Vaquero.etal2015}
{Vaquero} J.~M.,  {Kovaltsov} G.~A.,  {Usoskin} I.~G.,  {Carrasco} V.~M.~S.,
  {Gallego} M.~C.,  2015, \mn@doi [\aap] {10.1051/0004-6361/201525962}, 577,
  A71

\bibitem[\protect\citeauthoryear{{Watson}, {Fletcher}, {Dalla}  \&
  {Marshall}}{{Watson} et~al.}{2009}]{Watson.etal2009}
{Watson} F.,  {Fletcher} L.,  {Dalla} S.,   {Marshall} S.,  2009, \mn@doi
  [\solphys] {10.1007/s11207-009-9420-z}, \href
  {http://adsabs.harvard.edu/abs/2009SoPh..260....5W} {260, 5}

\bibitem[\protect\citeauthoryear{{Zolotova} \& {Ponyavin}}{{Zolotova} \&
  {Ponyavin}}{2015}]{Zolotova.Ponyavin2015}
{Zolotova} N.~V.,  {Ponyavin} D.~I.,  2015, \mn@doi [\apj]
  {10.1088/0004-637X/800/1/42}, 800, 42

\bibitem[\protect\citeauthoryear{{de La Lande} \& {Messier}}{{de La Lande} \&
  {Messier}}{1769}]{Delalande.Messier1769}
{de La Lande} M.,  {Messier} M.,  1769, \mn@doi [Philosophical Transactions of
  the Royal Society of London Series I] {10.1098/rstl.1769.0050}, \href
  {http://adsabs.harvard.edu/abs/1769RSPT...59..374D} {59, 374}

\makeatother
\end{thebibliography}

\bsp	
\label{lastpage}
\end{document}